# User-driven Privacy Enforcement for Cloud-based Services in the Internet of Things

Martin Henze*, Lars Hermerschmidt‡, Daniel Kerpen§, Roger Häußling§, Bernhard Rumpe‡, Klaus Wehrle*
*Communication and Distributed Systems, §Sociology of Technology and Organization, ‡Software Engineering
RWTH Aachen University, Germany
Email: {henze, wehrle}@comsys.rwth-aachen.de, {dkerpen, rhaeussling}@soziologie.rwth-aachen.de,
{hermerschmidt, rumpe}@se.rwth-aachen.de

*Abstract*—Internet of Things devices are envisioned to penetrate essentially all aspects of life, including homes and urban spaces, in use cases such as health care, assisted living, and smart cities. One often proposed solution for dealing with the massive amount of data collected by these devices and offering services on top of them is the federation of the Internet of Things and cloud computing. However, user acceptance of such systems is a critical factor that hinders the adoption of this promising approach due to severe privacy concerns. We present UPECSI, an approach for user-driven privacy enforcement for cloud-based services in the Internet of Things to address this critical factor. UPECSI enables enforcement of all privacy requirements of the user once her sensitive data leaves the border of her network, provides a novel approach for the integration of privacy functionality into the development process of cloud-based services, and offers the user an adaptable and transparent configuration of her privacy requirements. Hence, UPECSI demonstrates an approach for realizing user-accepted cloud services in the Internet of Things.

*Keywords*-Internet of Things, Cloud Computing, Privacy, Model-driven, User-acceptance, Services, Development

## I. Introduction

The Internet of Things (IoT) is the vision of interconnecting an incredible large amount of smart things (*IoT devices*), making its way into almost all aspects of everyday life. Typical examples for the envisioned deployment of IoT systems for end-users include, but are not limited to, pervasive health care, assisted living, and smart cities [1], [2]. IoT devices are often constrained in their computational and storage resources and might be powered by a battery. Additionally, they may be mobile or loose connectivity and hence cannot be assumed to be always available. To overcome these constraints, it seems natural to leverage the elastic and always available resources of the cloud [3]–[7]. Cloud solutions simplify storage and processing of collected data, usage of same data within several services, as well as combining data of several users and supporting user mobility without information fragmentation over several databases.

While necessary sensors and cloud solutions already exist today, user acceptance of such a comprehensive solution remains a critical factor [2], [8]–[10]. Especially, a user's privacy could be subject to a wide range of threats, notably for a widely deployed system. Hence, it is crucial to ensure users' data control – optionally being able to rate, e.g., in the context of pervasive health care, health higher than privacy in an emergency. This is only possible by anchoring privacy aspects within the system itself, oriented at users' demands.

As an IoT device usually is not custom made for a specific user, but a product for a whole group, privacy demands can not be hard-coded into the product. Instead, privacy choices have to be left to the user. This increases complexity during the design and development of IoT services. Additionally, they put a burden on the user who is often not aware of the (technical) consequences of her choices.

In order to overcome these issues, we present UPECSI, our envisioned solution for **U**ser-driven **P**rivacy **E**nforcement for **C**loud-based **S**ervices in the **I**oT. Specifically, with UPECSI, we make the following contributions: i) enforcement of a user's privacy requirements even if her sensitive data leaves the secured borders of her own IoT network, ii) a novel approach for integrating privacy functionality into the development of cloud-based services, and iii) an adaptable, flexible, simple, and transparent possibility for a user to configure her privacy settings. With these contributions, we lay the foundations for bringing the IoT and cloud computing together with a user-driven approach for privacy-enabled service development in a cloud-based system.

The remainder of this paper is structured as follows. First, in Section II we describe the use cases and network scenario of our approach. Then, we discuss privacy concerns of end-users and privacy considerations of service providers in such a scenario in Section III. This is followed by a discussion of related work in Section IV. In Section V, we present the main contribution of this paper, our envisioned UPECSI architecture. We conclude our paper in Section VI and give an outlook to the next steps of realizing UPECSI.

## II. Scenario

In this section, we provide an overview of our envisioned scenario, i.e., exemplary use cases as well as the network scenario from both a societal and technical point of view. We motivate and discuss the design of UPECSI in the contexts of assisted living and interactive assistance in public spaces.

### A. Use Cases

In the first use case, *assisted living*, we consider a 75-years-old widowed female retiree who appreciates living independently in her apartment. A number of unobtrusive sensors screen her vital signs and deliver information to the cloud for fast access, e.g., by family members and third parties such as doctors and health care providers. In the second context, we consider this lady being comfortable with seamlessly taking UPECSI with her on her portable and/or wearable devices, where it takes a proactive role to provide *assistance in public spaces*. Easing the management of data and services in both contexts, the user is assisted by UPECSI as an easy-to-use interface to the whole cloud-based system.

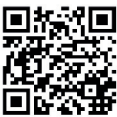

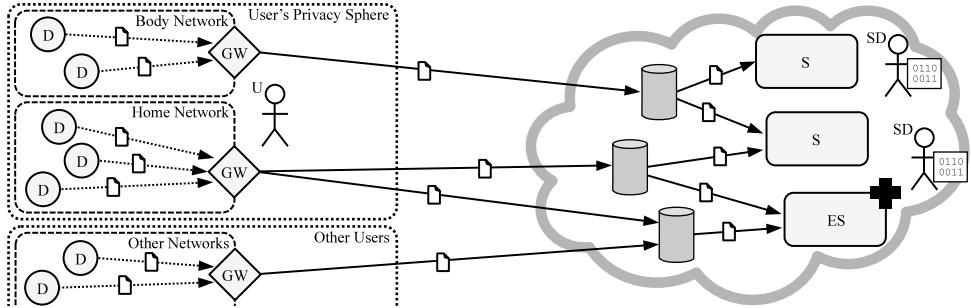

Figure 1. Each user (*U*) operates one or more networks of IoT devices (*D*) sending data to the cloud over a gateway (*GW*). The cloud stores data in different databases and provides it to user-authorized services (*S*), as well as emergency services (*ES*), which are provided by service developers (*SD*).

These use cases are influenced by social forces such as aging (western) societies and the therefore increasing role of technology in supporting formal and informal care/assistance. Furthermore, they do not only reflect general societal processes of differentiation, singularization, and changing gender roles, but are particularly prevalent when studying the changing social structure of older life [11]. Hence, they serve as adequate examples to derive design requirements for UPECSI to be user-accepted and provide us with a promising evaluation setting. However, the technical foundations of UPECSI are applicable to much broader contexts ranging from building management systems over intelligent transportation systems to smart manufacturing.

### B. Network Scenario

From a technical perspective, we depict our assumed network scenario in Figure 1. Each *user (U)* operates one or more networks of *IoT devices (D)*, e.g., a body network consisting of unobtrusive health care devices as well as a home network containing assisted living devices. Each of these networks is connected to the cloud over a dedicated *gateway (GW)* device. All devices (and hence networks) of a user form her *privacy sphere*, i.e., she has a certain trust into components within this sphere but does not want any potentially sensitive information collected in this sphere to be unrestrictedly available to third parties. Data that is sensed in one of the networks is forwarded via the gateway to the *cloud*. There, it is stored persistently and can be accessed by *services (S)* and *emergency services (ES)*. Note that we assume a large number of users to participate in this setting, leading to a multi-tenancy scenario. These services are offered by *service developers (SD)* who usually follow business interests and implement the functionality along with supporting functions such as billing or scaling with load.

### III. PRIVACY CONCERNS AND CONSIDERATIONS

In order to realize a scenario as described above, where potentially sensitive data from IoT devices is outsourced to the cloud, various privacy aspects have to be considered. We divide our discussion of these aspects in privacy concerns of end-users and privacy considerations of service providers.

### A. Privacy Concerns of End-users

Data sensed by IoT devices often contains sensitive information that others might be interested in [2]. For example, data from a pervasive health care and assisted living system might prove to be valuable for insurance companies in order to increase a person's fee or even deny a new contract. Additionally, not only the sensed data itself but also corresponding meta information (e.g., location or time) might be considered sensitive [12]. Thus, users often refrain from revealing the readings of their IoT devices to third parties.

Even more privacy concerns and issues arise when outsourcing this data to the cloud. For end-users, the major concern is the perceived loss of control over data if it is outsourced to the cloud [9], [10]. This mainly results from the fact that there is no control or at least transparency over the access to this data and, hence, data might be handed over to third parties or misused for unintended purposes [10]. Due to these concerns, end-users tend to refrain from using cloud-based services for (highly) sensitive data.

On a more general scale, an empirical study [13] showed that people expect a reasonable protection of their data, that legislation is adhered to, and that they are being informed about when and for what purpose their data is used. Additionally, privacy is not a fixed norm. Users are willing to trade-in privacy, e.g., for health in case of emergencies or quality of life in a nursing home [14]. This emphasizes the importance of including the user within privacy research, especially by providing transparency.

### B. Privacy Considerations of Service Providers

From a service provider's point of view, ignoring the previously depicted concerns can lead to undesired consequences, ranging from the nonacceptance of a service to costly lawsuits [2], [15]. Furthermore, especially when providing cloud services, legal restrictions, e.g., regarding the storage location and duration of data, have to be addressed [10], [15]. Particularly for legal requirements, service providers often need support from cloud providers [10].

In order to satisfy the user demand for transparency, the service provider has to lay out all uses of customer data. Nowadays, these descriptions are developed by lawyers, who have to interview the developers in order to get information about the processed data. As human communication is error prone, these descriptions are often formulated very defensive, so that lawsuits filed by users will not be successful even if the data handling deviates from the developer-intended behavior [16]. This lack of precise, up-to-date information about the data processing leads to uncertainty and non-transparency for the user.

## IV. RELATED WORK

Securely outsourcing data, which is collected by small devices, to the cloud has already been proposed by related work. For example, Lounis et al. [17] provide a secure cloud-based solution for health care data, while the SensorCloud project [3]–[5], [7], [18], [19] aims at providing a secure infrastructure for realizing services which operate on all kinds of sensor data. Although these approaches focus on security aspects, they do not address privacy concerns beyond providing confidentiality and access control.

In both domains, IoT and cloud computing, solutions for preserving privacy have been proposed. Examples for the IoT domain include the privacy-preserving computation of statistical values on health care data [20] and a context-aware system for assisted living and residential monitoring [21]. To ensure privacy in cloud computing, solutions ranging from annotating data with fine-grained privacy obligations [10] to using tamper-proof hardware components for realizing privacy-aware data storage and processing [22] have been proposed. Although these approaches present techniques for preserving privacy in the IoT and the cloud, we are not aware of a comprehensive approach for providing privacy in the combination of the two to support users and developers in using or, resp., designing services in a privacy-aware way.

In order to express privacy requirements, several approaches have been proposed [23]–[26]. Notably, the SPARCLE Workbench [27] transforms privacy policies from natural language to XACML [28] and back. This is shown to be useful for policy makers, auditors, and managers [29] but not for service developers, as these need a compact and comprehensive language, which adopts known concepts of their domain [30]. All these approaches can describe data handling policies, but only some enable user choice, and none is integrated with IoT/cloud service development.

Hence, related work provides valuable insights and building blocks, but in order to fully address the privacy concerns and considerations as discussed in Section III when combining IoT and cloud computing, additional effort is required. For this, it is crucial to focus on both the users and developers of an IoT/cloud service.

## V. SYSTEM OVERVIEW

To address these privacy concerns, we present the design of UPECSI, our system for user-driven privacy enforcement for cloud-based services in the IoT. The agreed requirements for privacy enforcement are notice, consent, self-determination, adequate security, and purposeful use [15]. Notice, consent, and self-determination can be achieved by interacting with the user and thus providing transparency, while adequate security and purposeful use can be achieved through technical measures.

In order to realize UPECSI, we identify the following three core aspects: i) *Privacy Enforcement Points*, which are situated on the network gateways and allow the user to enforce her security and privacy requirements beyond the networks she physically controls, ii) *Model-driven Privacy*, which eases the integration of privacy functionality into service development, and iii) *Interaction with the User* to provide transparency. Figure 2 gives a high-level overview of UPECSI by illustrating these core aspects and their interplay. First, the *Privacy Enforcement Points (PEP)* allow the user to protect access to her data and hence guarantee adequate security and consent. Second, using Model-driven Privacy, a service-specific *privacy policy (PP)* can automatically be derived. The service's realization and *data usage monitoring (M)* is audited by a *trusted third-party (TTP)* using *audit information (A)* to guarantee purposeful use and notice. The audited policy, together with a default *privacy configuration (PC)* recommended by a trusted third-party, is provided to the user on her *interface (I)*. There, to provide consent, she makes the final decision if and under which conditions she grants a service access to her data. In the following, we discuss the aspects of UPECSI in more detail.

### A. Privacy Enforcement Points

As we have shown in Section II, a network of IoT devices is connected to the cloud via a gateway. Additionally, a user might operate more than one network of IoT devices and hence more than one gateway. We observe, that these gateways are the distinct boundaries at which data leaves the control of the user. Hence, we introduce a new control element on each of these gateways, the *Privacy Enforcement Point (PEP)*. The PEP acts as a representative of the user and allows her to remain in control over security and privacy requirements regarding her data, even if it leaves the protected home network and is outsourced to the potentially untrusted cloud. In order to achieve this goal, the PEP has to fulfill three tasks, which we present in the following.

*1) Data Protection:* As a basis for privacy enforcement mechanisms, we need a security infrastructure, which offers the basic data protection functionalities access control and encryption. For this, we leverage the data protection and access control mechanisms of the SensorCloud security architecture [3], [4], [7]. Before any data is uploaded to the cloud, the PEP encrypts it using a symmetric data protection key. This protects the confidentiality of the data and prevents unauthorized access. In order to give an authorized cloud service access to the data, the PEP encrypts the data protection key with the public key of the cloud service and uploads it to the cloud. Then, the cloud service may decrypt the data protection key using its private key and, subsequently, decrypt the data it is authorized to access, too. In order to allow more fine-grained access control, data can be split into data fields that are encrypted separately, and data protection keys may be exchanged periodically, which allows to restrict access to certain periods of time.

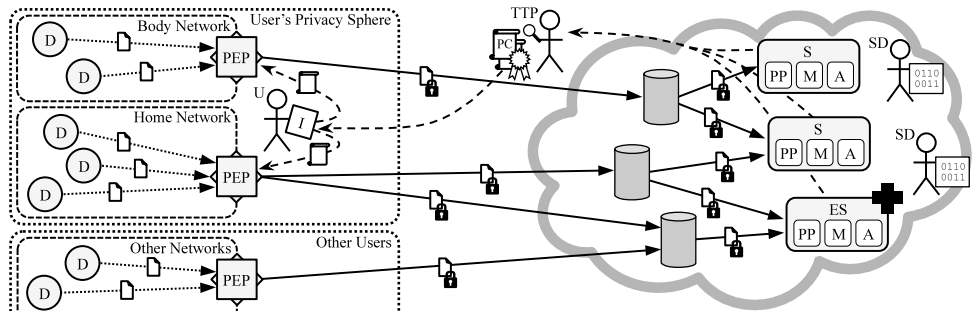

Figure 2. We introduce Privacy Enforcement Points (*PEP*), which encrypt data before sending it to the cloud. A Service provides a Privacy Policy (*PP*), Data Usage Monitoring (*M*), and Audit information (*A*) to a trusted third-party (*TPP*). The TPP reviews the information from the service and suggests default Privacy Configurations (*PC*) to the user, who reviews these on her Interface (*I*) and finally instructs her PEP to give the service access to her data.

*2) Annotation Enforcement:* In Section III, we have discussed that privacy requirements do not only consist of access to data but also impose restrictions on where data is transferred to. The PEP allows the user to annotate her data with such requirements [10], [31] and subsequently enforces them. The user might, e.g., specify that certain data might not leave her network at all, other data might only be forwarded to specified cloud offers, while a third group of data might be transferred to arbitrary cloud solutions. Thus, the PEP will, based on the annotation, decide whether data is allowed to leave the controlled network and if so, only forward the data to those parts of a cloud that have been authorized by the user to receive these data.

*3) Flexible Access Control:* As discussed in Section III, we have to account for a trade-off between privacy and other goods. In a medical emergency, e.g., the user would probably be willing to give up her privacy completely if this helps to save her life. However, the fixed and predefined access control mechanisms as presented in Section V-A1 prevent this. Hence, we have to support *Flexible Access Control* in UPECSI. For this, we extend the access control mechanisms to support an automatic authorization of cloud services based on internally and externally triggered events. Internally triggered events originate from one of the networks of the user and could, e.g., be triggered based on the measurements of an IoT device. Contrary, externally triggered events originate from outside the user's networks and are typically triggered by a trusted third party, e.g., a health professional or emergency center. Also, the combination of internally and externally triggered events is possible, e.g., an emergency doctor is only allowed access to the user's data, if his statement that the user is unconscious is consistent with readings from the user's vital signs monitoring device. The user defines upfront who should get access to her data resp. who makes that decision when an event is triggered. For example, she could choose that any emergency doctor gets access to her data in case of a medical emergency.

*B. Model-driven Privacy*

Considering privacy during service development requires additional effort from the service developers and potentially leads to more complexity. One approach to reduce complexity and ease the developer's job is to rise the abstraction of the programming language. The model-driven development (MDD) approach [32] proposes to use models instead of general purpose programming language code, which are then used to generate parts of the software [33]. The applicability of MDD has been shown in various contexts, ranging from cloud computing [18] to security and access control [34], [35]. However, all these approaches do not address privacy considerations.

As part of UPECSI, we introduce the Privacy Development Language (PDL) in order to allow cloud service developers to describe their privacy considerations while at the same time having object oriented programming elements like classes and methods. This way, we interconnect the service's data model and privacy description during service development. PDL can be used to automatically generate the implementation specific part of a service's privacy policy and data use monitoring. To realize PDL, we extend UML/P [36], a developer-friendly version of UML for modeling data structures and methods in a Java-like syntax.

Following Pearson's "top six" for software engineers [15], we identified *Privacy Policy*, *Monitoring Data Usage*, and *Audit Data Processing* as mandatory for services to meet users' privacy demands. These measures are highly dependent on the data structures and methods used within the

```
class Camera {
  <<use="Determine where to display image in house
       overview map.">>
  Room location;
  <<mandatory="Analysis.healthCritical">>
  <<optional="Analysis.getPersonalizedAd">>
  List<Person> recognizedPersons;
  VideoData stream;
}
class Analysis {
  <<use="Determine whether resident is alone and
       needs help.">>
  boolean healthCritical();
  <<use="Ad funded service">>
  <<notUsed="Fee is $1 per Month">>
  Ad getPersonalizedAd();
}
```

Listing 1. Example PDL domain model

service. We use the information modeled in PDL to automatically derive the privacy policy and monitoring of data usage. Following, we describe the generation of these parts from the PDL and the course of auditing data processing.

*1) Privacy Policy:* Commonly, a privacy policy is composed of service specific information and general liability disclaimers, required for every service and provided by the legal division. As noted before, the service-specific information is often very vague, due to the lack of more precise information from developers. With the PDL, information on the usage of user data is available, so that the service-specific part of the privacy policy can be automatically derived.

Listing 1 shows a PDL model written by a service developer. For each attribute, i.e., customer data originating from an IoT device, the service developer has to specify the use of customer data within the service with the use stereotype. Along with the class and attribute name, this information allows to express data usage in a detailed, human-readable way. We defer the discussion on how this information can be perceived by inexperienced users to Section V-C.

In order to allow the user to customize the functionality of a service to her privacy requirements, methods can be declared as mandatory or optional using the respective stereotype. Hence, each optional stereotype indicates a user-choice in the privacy policy. In case the service functionality changes if a method is disabled by the user, the consequences have to be specified in human-readable form in the notUsed stereotype. For example, with the method getPersonalizedAd() in Listing 1 the user has two options 1.) let the service use the list of recognized persons for an ad-funded service, or 2.) pay $1 per month. Hence, for each optional method, the user has to choose between the two mutually exclusive options use and notUsed.

*2) Monitoring Data Usage:* In order to provide transparency over the access to data, we propose a database layer monitoring data access in addition to the fine-grained access control as described in Section V-A3. The provider of the cloud platform logs every data access to an attribute with a use stereotype, i.e., all customer data. The log data is cryptographically protected to provide confidentiality and integrity, i.e., only the owner of the data being accessed can view the log, and no one can tamper with the log, as it has a public-key signature of a trusted third-party. This way, the user may, at any time, retrieve a detailed statement on which method accessed which parts of her data, at which time, and for which purpose.

*3) Audit Data Processing:* In order to take a profound privacy decision, the user needs to be sure that the individual parts of a service only use her data in the way described in use. However, a typical user (cf. Section II) does not have the ability to verify this by herself. Hence, a trusted auditor, e.g., the provider of the cloud platform or a certification authority, has to verify the data usage. As the data use description is connected to the implementing method, the auditor can efficiently perform a detailed review, as he can directly focus on the privacy relevant methods of the service and is aware of their expected behavior. Reviewing the application-specific flexible access control implementation, the auditor ensures that policies like the emergency doctor example in Section V-A3 in deed rely on the trusted third party. The provider of the cloud platform will only make those services available to the user, which have successfully been audited. Thus, the user is ensured that the service actually uses her data only as specified in the privacy policy.

*C. Interaction with the User*

In order to realize notice and consent, user interaction with UPECSI requires special consideration. When referring to the use cases of Section II, we must consider that UPECSI's approach to privacy may be applicable to users of different levels of privacy expertise, ranging from privacy experts able to make informed decisions up to absolutely IoT-privacy laymen – like our 75-years-old widowed female retiree.

Therefore, technically speaking, a user considered as an IoT-privacy expert may use her interface to read on the most detailed scale through the generated privacy policy of every service she uses. In this context, the decision which options to choose from the privacy policy or which data to submit to the service, respectively, is stored as a user's privacy configuration on the PEP. Contrary, default privacy settings at different layers of abstraction are provided by various kinds of trusted third-parties to support IoT-privacy laymen in their decisions about privacy configurations. Although using a default policy, the user optionally may still change individual settings in her privacy configuration. By offering such interaction possibilities, the user is continuously assured control and verifiability of her data regardless of her factual privacy-related technical expertise.

In order to establish trust in the data usage of the services, users rely on the verdict of an auditor. Users may decide in advance which information should be revealed to any medical personnel in case of emergencies. Here as well, there is an institution which guarantees that only certified personnel is granted access under specified emergency conditions.

Hence, from a sociology of information perspective, UPECSI rests upon context-specific norms constraining what information is collected, as well as with whom and under what conditions information is shared [37]. Thus, the system highlights not only a technical but also a social, i.e., situated and collective, context [38]. Finally, we consider this a useful approach to maximize transparency [39] of the whole system which will foster users' trust in UPECSI.

## VI. CONCLUSION AND OUTLOOK

In this paper, we discussed privacy concerns and considerations which hinder IoT and cloud computing federation from both end-users' and service providers' perspective. To tackle these critical concerns, we presented UPECSI as an approach for user-driven privacy enforcement for cloud-based services in the IoT. Our main focus rested on presenting its technical foundations with its three pillars privacy enforcement points, model-driven privacy, and user interaction. Its configurability on different layers of abstraction minimizes critical privacy concerns of different user groups and helps to increase user acceptance promoting UPECSI's adoption in further application areas.

In the future, we will further validate the feasibility of our proposed solution. For this, we are currently working on prototypes for the three pillars of UPECSI. To spur its development process, UPECSI's future technical build-up will be tied to feedback loops of our participatory design approach [5]. In doing so, we incorporate different interest groups such as end-users, service providers, experts, as well as other stakeholders and empower them as active members of technology development to increase user acceptance.

## ACKNOWLEDGMENT

This work has been funded by the Excellence Initiative of the German federal and state governments within the HumTec Project House at RWTH Aachen University.